\documentclass[aps,prl,reprint,superscriptaddress,footinbib]{revtex4-2}
\usepackage{amsmath, bm}
\usepackage{graphicx}

\begin{document}

\title{Resummation of Threshold Double Logarithms in 
Hadroproduction\\ of Heavy Quarkonium}

\author{Hee Sok Chung}
\email{neville@korea.ac.kr}
\altaffiliation[Present address:]{~Department of Mathematics and Physics,
Gangneung-Wonju National University, Gangneung 25457, Korea.}
\affiliation{Department of Physics, Korea University, Seoul 02841, Korea}

\author{U-Rae Kim}
\email{kim87@kma.ac.kr}
\affiliation{Department of Physics, Korea Military Academy, Seoul 01805, Korea}

\author{Jungil Lee}
\email{jungil@korea.ac.kr}
\affiliation{Department of Physics, Korea University, Seoul 02841, Korea}

\date{\today}

\begin{abstract}
We resum threshold double logarithms that appear in inclusive production of
heavy quarkonium. This resolves the catastrophic failure of fixed-order
perturbation theory where quarkonium cross sections at large transverse
momentum can turn negative due to large radiative corrections. We find that
resummation is imperative for describing measured prompt production rates of
$J/\psi$ at large transverse momentum. 
\end{abstract}

\maketitle

{\it Introduction---}Recent measurements of charmonium production rates at the LHC 
at very large transverse momentum~\cite{ATLAS:2023qnh} 
posed a serious challenge to the
phenomenology of heavy quarkonium production based on the 
nonrelativistic QCD (NRQCD) factorization formalism~\cite{Bodwin:1994jh}. 
This formalism describes the inclusive cross section of a heavy quarkonium in
terms of sums of products of perturbatively calculable short-distance
coefficients (SDCs) and nonperturbative long-distance matrix elements (LDMEs). 
Predictions based on SDCs computed at next-to-leading-order (NLO) 
accuracy in fixed-order (FO) perturbation theory fail to describe the measured
$J/\psi$ production rates at the LHC for transverse momentum much larger
than 100~GeV. Moreover, predictions based on FO calculations of
SDCs can yield unphysical, negative cross sections at large transverse 
momentum~\cite{Chung:2023ext, Chen:2023gsu}. 
It has been suggested that this problem arises from threshold logarithms, which
originate from singularities in radiative corrections near boundaries of phase
space~\cite{Chung:2023ext, Chen:2023gsu}. 
Hence, the catastrophic failure of FO perturbation theory may be
resolved by resumming the threshold logarithms to all orders in perturbation
theory, which has not yet been done in the NRQCD factorization formalism for
production of heavy quarkonium. 

In this Letter, we resum the threshold logarithms in the NRQCD factorization
formula for production of $J/\psi$, $\psi(2S)$, and $\chi_{cJ}$ to all orders
in perturbation theory. We work at the leading double logarithmic level, 
which produces the dominant singularities in the SDCs. We focus on the
large-transverse momentum region, where the effect of resummation is most
significant. By using the Grammer-Yennie approximation~\cite{Grammer:1973db}, 
we obtain expressions
for the singularities in the SDCs in terms of soft functions, which make
possible the resummation of threshold logarithms by exponentiation. 
The resummed SDCs that we obtain are free of singularities, so that we can 
ensure the positivity of heavy quarkonium cross sections at large $p_T$.  
We then compute $J/\psi$ production rates at large $p_T$ from $pp$ collisions
at $\sqrt{s} = 13$~TeV by using our resummed SDCs and find that we obtain
substantially improved descriptions of LHC data compared to FO calculations. 

{\it NRQCD factorization and fragmentation---}In the NRQCD factorization 
formalism, the inclusive cross section of a heavy
quarkonium ${\cal Q}$ at large transverse momentum $p_T$ is given
by~\cite{Bodwin:1994jh} 
\begin{equation}
\sigma_{{\cal Q}} = \sum_{\cal N} 
\sigma_{Q \bar Q ({\cal N})} \langle {\cal O}^{\cal Q}({\cal N}) \rangle, 
\end{equation}
where $\sigma_{Q \bar Q ({\cal N})}$ is the SDC 
that corresponds to the production rate of a heavy quark ($Q$) and a heavy 
antiquark ($\bar Q$) pair in a specific color and angular momentum state
${\cal N}$, and $\langle {\cal O}^{\cal Q}({\cal N}) \rangle$ is
the LDME that represents the probability for a
$Q \bar Q$ in the state $\cal N$ to evolve into a quarkonium ${\cal Q}$. 
The sum over $\cal N$ can be truncated at a chosen accuracy in the
nonrelativistic expansion; for ${\cal Q} = J/\psi$ or $\psi(2S)$, dominant
contributions come from $^3S_1^{[1]}$, $^3S_1^{[8]}$, $^1S_0^{[8]}$, and 
$^3P_J^{[8]}$ channels, while for ${\cal Q} = \chi_{cJ}$ the 
$^3P_J^{[1]}$ and $^3S_1^{[8]}$ channels appear at leading order in the 
nonrelativistic expansion. 
Because we are interested in the large-$p_T$ region, we focus on the 
leading large-$p_T$ behaviors of the SDCs which are given by
leading-power (LP) fragmentation~\cite{Collins:1981uw, Nayak:2005rt}
\begin{equation}
\label{eq:LPfrag}
\sigma_{Q \bar Q ({\cal N})}^{\rm LP} = \sum_{i=g,q,\bar{q}} 
\int_0^1 dz \, \hat{\sigma}_{i(K)} D_{i \to Q \bar Q ({\cal N})} (z), 
\end{equation}
where 
$\hat{\sigma}_{i (K)}$ is the production rate of a massless parton $i = g$, 
$q$, or $\bar{q}$ with momentum $K$, 
$D_{i \to Q \bar Q ({\cal N})} (z)$ is the fragmentation 
function (FF) for fragmentation of a parton $i$ into $Q \bar Q ({\cal N})$, 
and $z = P^+/K^+$ with $P$ the $Q \bar Q$ momentum and 
the $+$ direction defined along $P$ in the lab frame. 
The correction to Eq.~(\ref{eq:LPfrag}) is suppressed by $m^2/p_T^2$, 
with $m$ the heavy quark mass, and can be written in
terms of next-to-leading-power (NLP) fragmentation~\cite{Kang:2011zza,
Kang:2014tta, Ma:2014svb}. 
In large-$p_T$ hadroproduction, the cross section is dominated by gluon
fragmentation ($i=g$). 
Gluon FFs for ${\cal N}= {}^3S_1^{[8]}$, $^3P_J^{[8]}$, and
$^3P_J^{[1]}$ involve distributions that are singular at
$z=1$~\cite{Braaten:1996rp, Bodwin:2003wh, Bodwin:2012xc}. It is useful to
define the Mellin transform 
\begin{equation}
\label{eq:Mellin}
\tilde{D}_{i \to Q \bar Q ({\cal N})} (N) = \int_0^1 dz \, z^{N-1} 
D_{i \to Q \bar Q ({\cal N})} (z). 
\end{equation}
The inverse Mellin transform is singular at $z=1$ if 
$\tilde{D}_{i \to Q \bar Q ({\cal N})} (N)$ does not vanish as $N \to \infty$. 
At leading order in the strong coupling $\alpha_s$, 
$\tilde{D}^{\rm LO}_{g \to Q \bar Q (^3S_1^{[8]})} (N)$ is a nonzero constant
as $N \to \infty$, 
while $\tilde{D}^{\rm LO}_{g \to Q \bar Q (^3P_J^{[8]})} (N)$ and 
$\tilde{D}^{\rm LO}_{g \to Q \bar Q (^3P_J^{[1]})} (N)$ diverge like $\log N$, 
which imply that these FFs involve singular
distributions such as delta functions and plus
distributions~\cite{Braaten:1996rp, Bodwin:2003wh, Bodwin:2012xc}. 
Moreover, these singularities are exacerbated by radiative corrections; 
the NLO corrections involve double logarithms in $N$ that are proportional to 
$\alpha_s \log^2 N$ times the leading-order result, 
which correspond to 
$[\log (1-z)/(1-z)]_+$ for $^3S_1^{[8]}$ and 
$[\log^2 (1-z)/(1-z)]_+$ for $^3P_J^{[8]}$ and 
$^3P_J^{[1]}$~\cite{Braaten:2000pc, Lee:2005jw, Ma:2013yla, Zhang:2020atv}. 
Because these 
logarithms are associated with singularities at 
$z=1$, we refer to them as threshold logarithms. 
We find that these double logarithmic terms 
almost completely reproduce the strong logarithmic dependences on
$p_T$ of the NLO corrections to $\sigma_{Q \bar Q ({\cal N})}$.
This strongly suggests that
resumming threshold logarithms is the key for resolving the negative
cross section problem.

{\it Soft approximation---}In order to resum the singularities in the FFs, we
obtain an expression for the FFs valid near $z=1$ by using the Grammer-Yennie
approximation for soft-gluon attachments to the $Q$ and $\bar Q$ produced from
the fragmenting gluon~\cite{Grammer:1973db, Collins:1981uk, Nayak:2005rt}. 
We obtain 
\begin{align}
\label{eq:softapprox}
D_{g \to Q \bar Q}^{\rm soft} (z)
&= 
2 M (-g_{\mu \nu} ) 
C_{\rm frag} 
\left|\frac{i}{K^2+i \varepsilon}\right|^2 
\nonumber \\ & \quad \times 
\left( g^{\mu \alpha} - \frac{K^\mu n^\alpha}{K^+} \right)
\left( g^{\nu \beta} - \frac{K^\nu n^{\beta}}{K^+} \right)
\nonumber \\ & \quad \times 
\langle 0 | 
\bar{T} [ {\cal A}_{\rm soft}^{\beta,c} \Phi_n^{bc} ]^\dag
\delta^+_z 
T [{\cal A}_{\rm soft}^{\alpha,a} \Phi_n^{ba} ]
| 0 \rangle,
\end{align}
where $M = \sqrt{P^2}$, 
$C_{\rm frag} = z^{d-3} K^+/[2 \pi (N_c^2-1) (d-2)]$, 
$d=4-2 \epsilon$ is the number of spacetime dimensions, 
$N_c$ is the number of colors, 
$n$ is a lightlike vector defined through $K^+ = n \cdot K$, 
$|0\rangle$ is the QCD vacuum, 
$\Phi_k \equiv 
\Phi_k (\infty,0) = {\cal P} \exp [ -i g \int_0^\infty d
\lambda \, k \cdot A^{\rm adj} (k \lambda) ]$ is a Wilson line in the
adjoint representation defined along a vector $k$, 
with $\cal P$ the path ordering, $g$ the strong coupling, and $A$ the gauge field, $T$ and $\bar T$ are time and antitime
orderings, respectively, and we use the shorthand $\delta^+_z 
\equiv 2 \pi \delta[n \cdot \hat p - (1-z) P^+]$, where $\hat p$ is an operator
that reads off the momentum of the operator to the right.
The operator ${\cal A}_{\rm soft}^{\alpha,a}$ represents 
an arbitrary number of soft-gluon attachments onto the $Q$ and $\bar Q$ 
lines in the soft approximation, and is given by 
\begin{equation}
\label{eq:softOP}
{\cal A}_{\rm soft}^{\alpha,a} = 
\bar{u} (p_1) W_{p_1} (\infty,0) (-i g \gamma^\alpha T^a) 
W_{p_2}^\dag (\infty,0) v (p_2), 
\end{equation}
where $p_1$ and $p_2$ are the momenta of the $Q$ and $\bar Q$, respectively, 
and $W_{k} (t',t) = {\cal P} \exp[-i g \int_t^{t'} d \lambda \, k \cdot A (k
\lambda)]$ is a Wilson line in the fundamental representation defined along a
vector $k$. 
Note that the operators on the right and left of the $\delta^+_z$ are always 
time ordered and antitime ordered, respectively. Hereafter we omit the 
time and antitime ordering symbols in expressions involving $\delta^+_z$.

In order to obtain expressions for specific ${\cal N}$, we first
expand in powers of the relative momentum $q \equiv (p_1 - p_2)/2$ and then
project onto specific color and angular momentum states. 
In the $^3S_1^{[8]}$ case, we set $q = 0$ in Eq.~(\ref{eq:softapprox})
to obtain 
\begin{align}
\label{eq:Dsoft3S1}
D_{g \to Q \bar Q(^3S_1^{[8]})}^{\rm soft} (z)
&= \frac{C_{\rm frag} (d-2) g^2}{4 m^3 (d-1) (N_c^2-1)} S_{^3S_1^{[8]}} (z), 
\end{align}
where $S_{^3S_1^{[8]}} (z)$ is the soft function defined by 
\begin{align}
\label{eq:soft3S1}
S_{^3S_1^{[8]}} (z) &=
\langle 0 | [ \Phi_p^{ca} \Phi_n^{ba} ]^\dag 
\delta^+_z
\Phi_p^{cd} \Phi_n^{bd} | 0 \rangle,
\end{align}
with $p \equiv (p_1+p_2)/2 = P/2$. 
The factor $(d-1) (N_c^2-1)$ in the denominator comes from the normalization of
the LDME. 
For the $^3P^{[8]}$ case, we expand Eq.~(\ref{eq:softapprox}) 
to linear order in $q$. 
The expansion of the Wilson line
can be carried out by using a straightforward generalization
of Polyakov's identity~\cite{Polyakov:1980ca, Nayak:2005rt}. 
We obtain 
\begin{align}
\label{eq:Dsoft3P8}
D_{g \to Q \bar Q(^3P_J^{[8]})}^{\rm soft} (z)
&= - \frac{C_{\rm frag} (d-2) g^4
S_{^3P^{[8]}} (z) 
}{4 m^3 (d-1)^2 (N_c^2-1)} ,
\end{align}
where the soft function $S_{^3P^{[8]}} (z)$ is given by 
\begin{align}
\label{eq:soft3P8}
S_{^3P^{[8]}} (z) &=
\langle 0 | [ {\cal W}^{yx}_{\alpha} ]^\dag
\delta^+ _z
{\cal W}_\beta^{y x} | 0 \rangle g^{\alpha \beta},
\end{align}
with 
\begin{align}
\label{eq:W3P8}
{\cal W}^{yx}_\beta &= \int_0^\infty d \lambda \, \lambda
\Phi_p^{yc} (\infty, \lambda) p^\mu G^b_{\mu \beta} (p \lambda) d^{bcd} 
\nonumber \\ & \quad \hspace{18ex} \times 
\Phi_p^{da} (\lambda,0) \Phi_n^{xa}, 
\end{align}
where $G_{\mu \nu} = \partial_\mu A_\nu - \partial_\nu A_\mu +i g [A_\mu,
A_\nu]$ is the QCD field-strength tensor. 
Finally, for the $^3P_J^{[1]}$ case, we have 
\begin{align}
\label{eq:Dsoft3P1}
D_{g \to Q \bar Q(^3P_J^{[1]})}^{\rm soft} (z)
&= - \frac{C_{\rm frag} (d-2) g^4}{4 N_c^2 m^3 (d-1)^2} 
\frac{9}{(2 J+1)}
\nonumber \\ & \quad \times \!
\left[ c_J S_{^3P^{[1]}} (z) 
+ c_J^{TT} S^{TT}_{^3P^{[1]}} (z) \right], 
\end{align}
where $c_0 = (d-1)^{-2}$, $c_1 = (d-2)/[2 (d-1)]$, $c_2 = (d-2) (d+1)/[2
(d-1)^2]$, $c_0^{TT} = [(d-1) (d-2)]^{-1}$, $c_1^{TT} = -[2 (d-2)]^{-1}$, 
$c_2^{TT} = (d-3)/[2 (d-1) (d-2)]$, 
and the soft functions are given by 
\begin{subequations}
\begin{align}
\label{eq:soft3P1}
S_{^3P^{[1]}} (z) &=
\langle 0 | [ {\cal \bar W}^{b}_{\alpha} ]^\dag
\delta^+ _z
{\cal \bar W}_\beta^{b} | 0 \rangle g^{\alpha \beta},
\\
S_{^3P^{[1]}}^{TT} (z) &=
\langle 0 | [ {\cal \bar W}^{b}_{\alpha}]^\dag
\delta^+ _z
{\cal \bar W}_\beta^{b} | 0 \rangle 
\nonumber \\ & \quad \times 
\left[ \frac{p^2 n^\alpha n^{\beta}}{(n \cdot p)^2} 
+ \frac{g^{\alpha \beta}}{d-1} \right], 
\end{align}
\end{subequations}
with 
\begin{align}
\label{eq:W3P1}
{\cal \bar W}^{b}_\beta &= \int_0^\infty d \lambda \, \lambda
p^\mu G^d_{\mu \beta} (p \lambda) 
\Phi_p^{da} (\lambda,0) \Phi_n^{ba}.
\end{align}
We use the definitions of the color-singlet LDMEs in Ref.~\cite{Bodwin:1994jh}, 
which differ from Refs.~\cite{Zhang:2017xoj, Zhang:2020atv} by a factor of $2
N_c$.  The $S_{^3P^{[1]}}^{TT} (z)$ term comes from the anisotropic
contribution that arises from projecting onto specific $J$. As we will see
later, 
$S_{^3P^{[1]}}^{TT} (z)$
will not produce double logarithmic
contributions and can be neglected in this work. 

The results in Eqs.~(\ref{eq:Dsoft3S1}), (\ref{eq:Dsoft3P8}), and 
(\ref{eq:Dsoft3P1}) reproduce the leading singularities of the FFs at $z = 1$,
which are contained in the soft functions. We can verify this at LO by
computing the ${\cal S}_{\cal N}$ at leading nonvanishing order: 
\begin{subequations}
\begin{align}
\label{eq:softLO}
S_{^3S_1^{[8]}}^{\rm LO} (z) &= \frac{2 \pi (N_c^2-1)}{P^+} \delta(1-z),
\\
S_{^3P^{[8]}}^{\rm LO} (z) &= - 
\frac{(d-2) 4 B_F (N_c^2-1)
\Gamma(1+\epsilon)}{2 \pi^{1-\epsilon} m^2 
P^+ (1-z)^{1+2 \epsilon}}, 
\\
S_{^3P^{[1]}}^{\rm LO} (z) &= -
\frac{(d-2) (N_c^2-1)
\Gamma(1+\epsilon)}{2 \pi^{1-\epsilon} m^2
P^+ 
(1-z)^{1+2 \epsilon}}, 
\\
S_{^3P^{[1]}}^{TT, \, {\rm LO}} (z) &= 
\frac{ (N_c^2-1) \epsilon(1- \epsilon) (1-2 \epsilon) \Gamma(1+\epsilon)}
{3 (3-2 \epsilon) \pi^{1-\epsilon} m^2 
P^+ (1-z)^{1+2 \epsilon}}, 
\end{align}
\end{subequations}
where $B_F = (N_c^2-4)/(4 N_c)$. 
Note that $S_{^3P^{[8]}}^{\rm LO}$, $S_{^3P^{[1]}}^{\rm LO}$, and 
$S_{^3P^{[1]}}^{TT, {\rm LO}}$ come from diagrams where a single gluon is
exchanged between the two field-strength tensors.  By using the identity 
\begin{align}
\label{eq:plusfunc}
\frac{1}{(1-z)^{1+n \epsilon}} 
= - \frac{1}{n \epsilon_{\rm IR}} \delta(1-z) + 
\left[ \frac{1}{(1-z)^{1+n \epsilon}} \right]_+, 
\end{align}
it is easy to show that the $D_{g \to Q \bar Q({\cal N})}^{\rm soft}(z)$
reproduce the singular distributions in the LO FFs~\cite{Braaten:1996rp, Bodwin:2003wh, Bodwin:2012xc}. 
Note that $S_{^3P^{[1]}}^{TT,\,{\rm LO}}$ 
does
not produce singular distributions because it contains an explicit factor of
$\epsilon$.

{\it Radiative corrections to soft functions---}We now compute the double logarithms in the soft functions at NLO, which come
from contributions involving double poles in $\epsilon$. 
We first consider $S_{^3S_1^{[8]}}(z)$. By explicit calculation we can show 
that in Feynman gauge, the double logarithms come from NLO diagrams where an
additional gluon is exchanged between the Wilson lines $\Phi_p$ and 
$\Phi_n$. The result is
\begin{align}
\label{eq:soft3S1diags}
S_{^3S_1^{[8]}}^{\rm NLO}(z) &=
\frac{2 \alpha_s C_A (N_c^2-1)}{P^+ (1-z)^{1+2 \epsilon}} 
\left( \frac{1}{\epsilon_{\rm UV}} + O(\epsilon^0) \right)
\nonumber \\ & \quad 
- \frac{\alpha_s C_A (N_c^2-1) }{P^+} \delta(1-z) 
\frac{1}{\epsilon_{\rm UV}} 
\nonumber \\ & \quad \quad \times 
\left( \frac{1}{\epsilon_{\rm UV}} - \frac{1}{\epsilon_{\rm IR}} \right)
+ \cdots,
\end{align}
where $C_A = N_c$ and 
we neglect any contributions that do not produce double logarithms. 
The contribution in the first line comes from the real diagram where the gluon
crosses the cut, while the remaining terms come from the virtual diagram. 
By using the identity in Eq.~(\ref{eq:plusfunc}), we obtain the expression for 
the double logarithmic correction to $S_{^3S_1^{[8]}}$ at NLO given by 
\begin{align}
\label{eq:soft3S1NLO}
S_{^3S_1^{[8]}}^{\rm NLO}(z) &=
\frac{2 \pi (N_c^2-1)}{P^+} 
\frac{\alpha_s C_A}{\pi} \bigg\{ \frac{-\delta(1-z)}{2 \epsilon_{\rm UV}^2}
+ \frac{1}{\epsilon_{\rm UV} (1-z)_+} 
\nonumber \\ & \quad 
+ \left[ \frac{-2 \log(1-z)}{1-z} \right]_+ + \cdots \bigg\}.
\end{align}
This reproduces the double 
logarithmic term $[\log(1-z)/(1-z)]_+$ in the $^3S_1^{[8]}$ FF at
NLO~\cite{Braaten:2000pc, Lee:2005jw, Ma:2013yla}. 

Similarly to the $^3S_1^{[8]}$ case, double logarithmic corrections to the 
$^3P^{[8]}$ and $^3P^{[1]}$ soft functions arise only from planar diagrams
where an additional gluon is exchanged between the 
Wilson lines $\Phi_p(\lambda,0)$ and $\Phi_n$.
The result is
\begin{subequations}
\begin{align}
\label{eq:soft3P8NLO}
S_{^3P^{[8]}}^{\rm NLO}(z) &=
\frac{\alpha_s C_A}{\pi} 
\frac{4 B_F (N_c^2-1) \epsilon_{\rm UV}^{-2}}{2 \pi m^2 [P^+ (1-z)]^{1+4
\epsilon}} 
+ \cdots, 
\\
\label{eq:soft3P1NLO}
S_{^3P^{[1]}}^{\rm NLO}(z) &=
\frac{\alpha_s C_A}{\pi}
\frac{(N_c^2-1)  \epsilon_{\rm UV}^{-2}}{2 \pi m^2 [P^+ (1-z)]^{1+4
\epsilon}} 
+ \cdots,
\end{align}
\end{subequations}
where we neglect any contribution that do not produce any double logarithmic
corrections. 
Again, by using Eq.~(\ref{eq:plusfunc}) we can check that the double
logarithmic term $[\log^2(1-z)/(1-z)]_+$ agrees with the explicit NLO 
calculation of the FFs in Ref.~\cite{Zhang:2020atv}. Similarly to the LO case,
$S_{^3P^{[1]}}^{TT}$ does not produce double
logarithmic singularities at NLO because it contains an explicit factor of
$\epsilon$. 

{\it Resummation---}It is straightforward to resum the double logarithmic
corrections to all orders in perturbation theory, by using the fact that they
arise from planar diagrams that can be exponentiated
in Mellin space~\cite{Laenen:2000ij}: 
\begin{equation}
\label{eq:softresum}
\tilde S_{\cal N}^{\rm resum} (N) =
\exp\left[ J_{\cal N}^N \right]
\tilde S_{\cal N}^{\rm LO} (N), 
\end{equation}
where $J_{\cal N}^N$ is given at leading double logarithmic level by 
\begin{subequations}
\begin{align}
\label{eq:softresumexponent}
J_{^3S_1^{[8]}}^N &= \frac{\alpha_s C_A}{\pi} 
 \int_0^1 dz \, z^{N-1} \left[ \frac{-2 \log(1-z)}{1-z} \right]_+, 
\\
J_{^3P^{[8]}}^ N&= 
J_{^3P^{[1]}}^N = 
\frac{4}{3} 
J_{^3S_1^{[8]}}^N. 
\end{align}
\end{subequations}
By expanding Eq.~(\ref{eq:softresum}) in powers of 
$\alpha_s$ we reproduce the double logarithmic corrections at NLO. 
The resummed expressions for the FFs can be obtained in the same way. 
At NLO accuracy, the resummed FFs can be written as 
\begin{align}
\label{eq:Dresum}
\tilde D_{g \to Q \bar Q ({\cal N})}^{\rm resum} (N) 
&=
\exp\left[ J_{\cal N}^N \right]
\times 
\bigg(
\tilde D_{g \to Q \bar Q ({\cal N})}^{\rm FO} (N) 
\nonumber \\ & \quad \quad
- J_{\cal N}^N 
\tilde D_{g \to Q \bar Q ({\cal N})}^{\rm LO} (N)
\bigg), 
\end{align}
where the last term in the parentheses subtracts the double logarithmic
correction term in the $\tilde D_{g \to Q \bar Q({\cal N})}^{\rm FO} (N)$ 
at NLO accuracy to avoid double counting. 
Note that, because the factor $\exp [J_{\cal N}^N]$ vanishes as $N \to \infty$ 
faster than any power of $N$, the inverse Mellin transform yields regular
functions in $z$ that vanish at $z=1$. 
The resummed expression for $^3S_1^{[8]}$ agrees with the
calculation in the soft-gluon factorization formalism in
Refs.~\cite{Chen:2023gsu, Chen:2021hzo} at double logarithmic level. 
The resummed results for $^3P_J^{[8]}$ and $^3P_J^{[1]}$ are new. 

We note that different expressions for the resummed FFs are possible that are
equivalent to Eq.~(\ref{eq:Dresum}) at the current accuracy. For
example, the NLO terms in Eq.~(\ref{eq:Dresum})
may be expanded in powers of $\alpha_s$, or an expression based on evolution
equations may be used~\cite{Chen:2021hzo}. We find that these alternative
expressions lead to cross sections that differ by less than 15\%, which is not
significant compared to other uncertainties in the theory of heavy quarkonium
production.

\begin{figure}[t]
\includegraphics[width=\columnwidth]{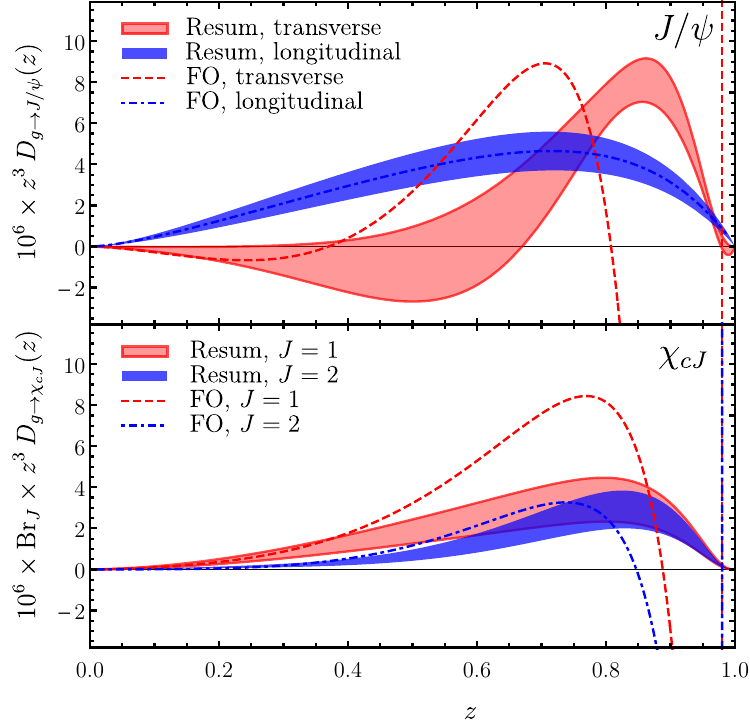}%
\caption{\label{fig:FFs} 
Gluon FFs with resummed threshold double logarithms 
times $z^3$ for production of $J/\psi$ (top) and $\chi_{cJ}$ (bottom) 
for $J=1$ and 2. Central values of FO results are also shown for comparison. 
${\rm Br}_J \equiv {\rm Br}_{\chi_{cJ} \to J/\psi + \gamma}$ is the branching
fraction for decays of $\chi_{cJ}$ into $J/\psi+\gamma$. 
}
\end{figure}

{\it Numerical results---}We now show the numerical results for the resummed FFs. 
We work with FFs for production of quarkonium ${\cal Q}$
written in terms of $D_{i \to Q \bar Q ({\cal N})} (z)$ as
\begin{align}
\label{eq:FFNRQCD}
D_{g \to {\cal Q}} (z)
&= 
\sum_{\cal N} 
D_{g \to Q \bar Q ({\cal N})} (z)
\langle {\cal O}^{\cal Q} ({\cal N}) \rangle, 
\end{align}
where the sum is over ${\cal N} = {}^3S_1^{[1]}$, $^3S_1^{[8]}$, $^1S_0^{[8]}$,
and $^3P_J^{[8]}$ for ${\cal Q} = J/\psi$ or $\psi(2S)$, 
and ${\cal N} = {}^3P_J^{[1]}$ and $^3S_1^{[8]}$ for ${\cal Q} = \chi_{cJ}$. 
Note that we include the contributions from ${}^3S_1^{[1]}$ and 
$^1S_0^{[8]}$, which do not contain singularities and are not affected
by resummation at the current level of accuracy~\cite{Braaten:1993rw,
Braaten:1995cj, Bodwin:2012xc, Zhang:2017xoj}.  
For consistency, we compute the FO FFs to order $\alpha_s^2$, 
except for the ${}^3S_1^{[1]}$ FF, which we compute at leading
nonvanishing order  
[$O(\alpha_s^3)$]~\cite{Braaten:1993rw, Braaten:1995cj, Zhang:2017xoj}. 
Because we are interested in the large-$p_T$ region, 
we evolve the FFs to the $\overline{\rm MS}$ scale 50~GeV from 
3~GeV by using the
Dokshitzer-Gribov-Lipatov-Altarelli-Parisi (DGLAP) evolution
equation~\cite{Gribov:1972ri, Lipatov:1974qm, Dokshitzer:1977sg,
Altarelli:1977zs} at leading logarithmic (LL) accuracy. 
We take the $J/\psi$ LDMEs determined in Ref.~\cite{Brambilla:2022ayc} in the
large-$p_T$ region, 
and we use the $\chi_{cJ}$ LDMEs from Ref.~\cite{Brambilla:2021abf}. 
The LDMEs are renormalized in the $\overline{\rm MS}$ scheme at the scale $m$. 
In order to compensate for the fact that resummation enhances the relative 
size of the $^3P^{[8]}$ SDCs compared to $^3S_1^{[8]}$ by about 10\%,
from which the $^3P^{[8]}$ LDME was determined in
Ref.~\cite{Brambilla:2022ayc}, we reduce the central value of the $^3P^{[8]}$
LDME by 10\%.  That is, we use 
$\langle O^{J/\psi} (^3S_1^{[1]}) \rangle = 1.18 \pm 0.35$~GeV$^3$, 
$\langle O^{J/\psi}(^3S_1^{[8]})\rangle=(1.40\pm0.42)\times10^{-2}$~GeV$^3$, 
$\langle O^{J/\psi}(^1S_0^{[8]})\rangle=(-0.63\pm3.22)\times10^{-2}$~GeV$^3$, 
$\langle O^{J/\psi}(^3P_0^{[8]})\rangle=(5.25\pm1.86)\times10^{-2}$~GeV$^5$, 
$\langle O^{\chi_{c0}}(^3P_0^{[1]})\rangle=(8.16\pm2.45)\times10^{-2}$~GeV$^5$, 
and
$\langle O^{\chi_{c0}}(^3S_1^{[8]})\rangle=(1.57\pm0.47)\times10^{-3}$~GeV$^3$. 
Note that due to the universality relations in the LDMEs obtained in
Refs.~\cite{Brambilla:2022rjd, Brambilla:2022ayc}, the 
$\psi(2S)$ FF can be obtained by uniformly rescaling the $J/\psi$ FF.
We display the gluon FFs for transversely and longitudinally polarized $J/\psi$ 
in Fig.~\ref{fig:FFs}. 
We show the FFs multiplied by $z^3$, because when computing $p_T$-differential
cross sections using Eq.~(\ref{eq:LPfrag}), 
the $\hat{\sigma}_{i}$ behave approximately like $z^3$. 
For comparison, we also show results for FO FFs, which involve
singularities at $z=1$ that cannot be displayed like regular functions. 
We see that the resummed $J/\psi$ FFs are 
smooth functions of $z$, and they are positive or at least consistent with zero
within uncertainties for all $0 < z < 1$, which ensures the positivity of
$J/\psi$ production rates. In contrast, the transversely polarized FF in FO
perturbation theory rapidly changes sign near $z=1$. The longitudinal
$J/\psi$ FF is unaffected by resummation, because it is free of singularities
at the current level of accuracy~\cite{Braaten:2000pc, Ma:2015yka}. 
The results for the resummed FFs lead to an
estimate of the polarization of $J/\psi$ and $\psi(2S)$ at $p_T = 100$~GeV 
given by $-0.25 \lesssim \lambda_\theta \lesssim +0.15$ in the helicity frame
at midrapidity, which is compatible with previous
estimates~\cite{Brambilla:2022rjd, Brambilla:2022ayc} but smaller than
recent CMS measurements at large $p_T$~\cite{CMS:2024igk}. 
While in the resummed case we expect $\lambda_\theta$ to stay almost constant 
as $p_T$ increases, in the FO calculation the parameter may unphysically drop 
below $-1$ when the cross section turns negative, because the 
longitudinal contribution is unaffected by 
threshold logarithms.
Similarly, we show the FFs for $\chi_{c1}$ and $\chi_{c2}$ in 
Fig.~\ref{fig:FFs}, scaled by the branching fractions into $J/\psi+\gamma$ 
from PDG~\cite{ParticleDataGroup:2024}. 
Just like the $J/\psi$ case, the resummed $\chi_{cJ}$ FFs are smooth functions 
of $z$ that are positive for the whole range of $z$, unlike the FO calculations
which change sign rapidly near $z=1$. 

\begin{figure}[t]
\includegraphics[width=\columnwidth]{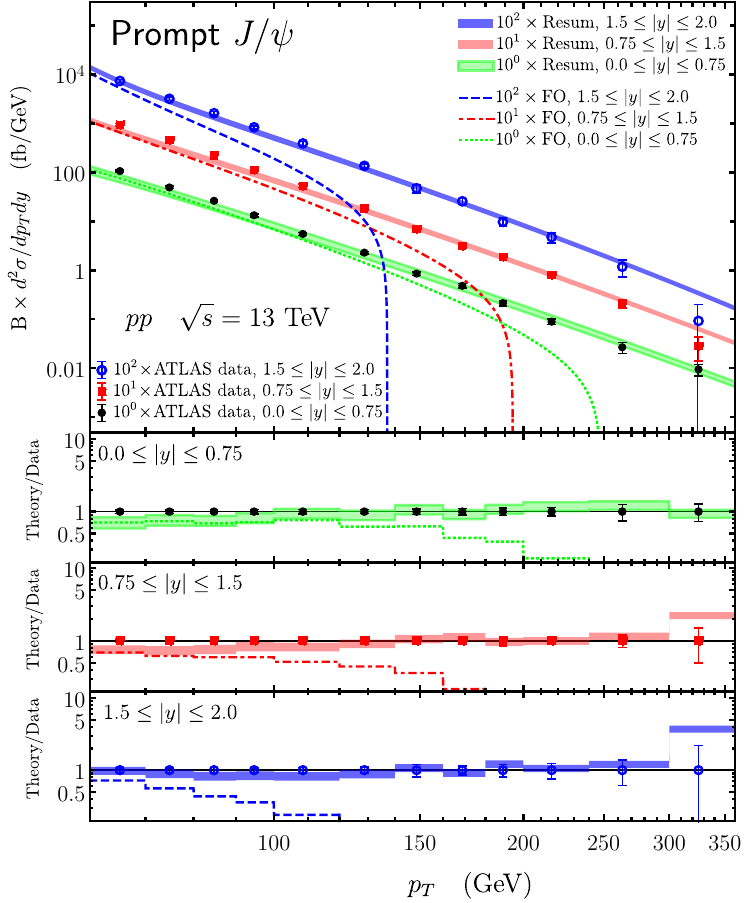}%
\caption{\label{fig:ATLAS}
Prompt $J/\psi$ production rates from $pp$ collisions at $\sqrt{s}=13$~TeV 
computed from resummed SDCs compared to ATLAS data. 
Central values of FO NLO results are shown for comparison. 
${\rm B} \equiv {\rm Br}_{J/\psi \to \mu^+ \mu^-}$ is the
$J/\psi$ dimuon branching fraction. 
}
\end{figure}

Finally, we compute the prompt $J/\psi$ production rates at the
$\sqrt{s}=13$~TeV LHC to compare with ATLAS measurements in
Ref.~\cite{ATLAS:2023qnh}. 
We use the method used in Ref.~\cite{Bodwin:2015iua} to compute the cross
sections in LP fragmentation, 
except that we use the gluon FFs including threshold
double logarithms and DGLAP logarithms resummed to LL accuracy to compute the LP
contribution. We also include the
contributions from light-quark FFs, which are not affected by resummation at
the current level of accuracy. We include feeddown contributions from decays of
$\chi_{cJ}$ and $\psi(2S)$, with branching fractions taken from
PDG~\cite{ParticleDataGroup:2024}. 
We include the NLP contributions we obtain from 
FO SDCs 
computed at NLO from the {\sc FDCHQHP} package~\cite{Wan:2014vka}. 
We find that the NLP contributions amount to about 10\% at $p_T = 60$~GeV 
and diminish to less than 1\% for $p_T$ larger than 100~GeV. 
We show the large-$p_T$ cross sections computed from the resummed FFs in
Fig.~\ref{fig:ATLAS} compared to ATLAS data~\cite{ATLAS:2023qnh}. 
The resummed results are in fair
agreement with data in the large-$p_T$ region. 
In contrast, the results from FO SDCs~\footnote{
While we use {\sc FDCHQHP} to compute the FO SDCs, for values of $p_T$ larger than
100~GeV, we compute the FO SDCs from LP fragmentation by using FO FFs. The
agreement between LP fragmentation and FO SDCs at large $p_T$ has been verified
in Ref.~\cite{Bodwin:2015iua}.} 
shown in Fig.~\ref{fig:ATLAS} 
fall below measured data and turn negative at large $p_T$. 
This shows that resummation of threshold logarithms is absolutely necessary in
order to describe heavy quarkonium production rates at very large transverse
momentum. 

\begin{figure}[t]
\includegraphics[width=\columnwidth]{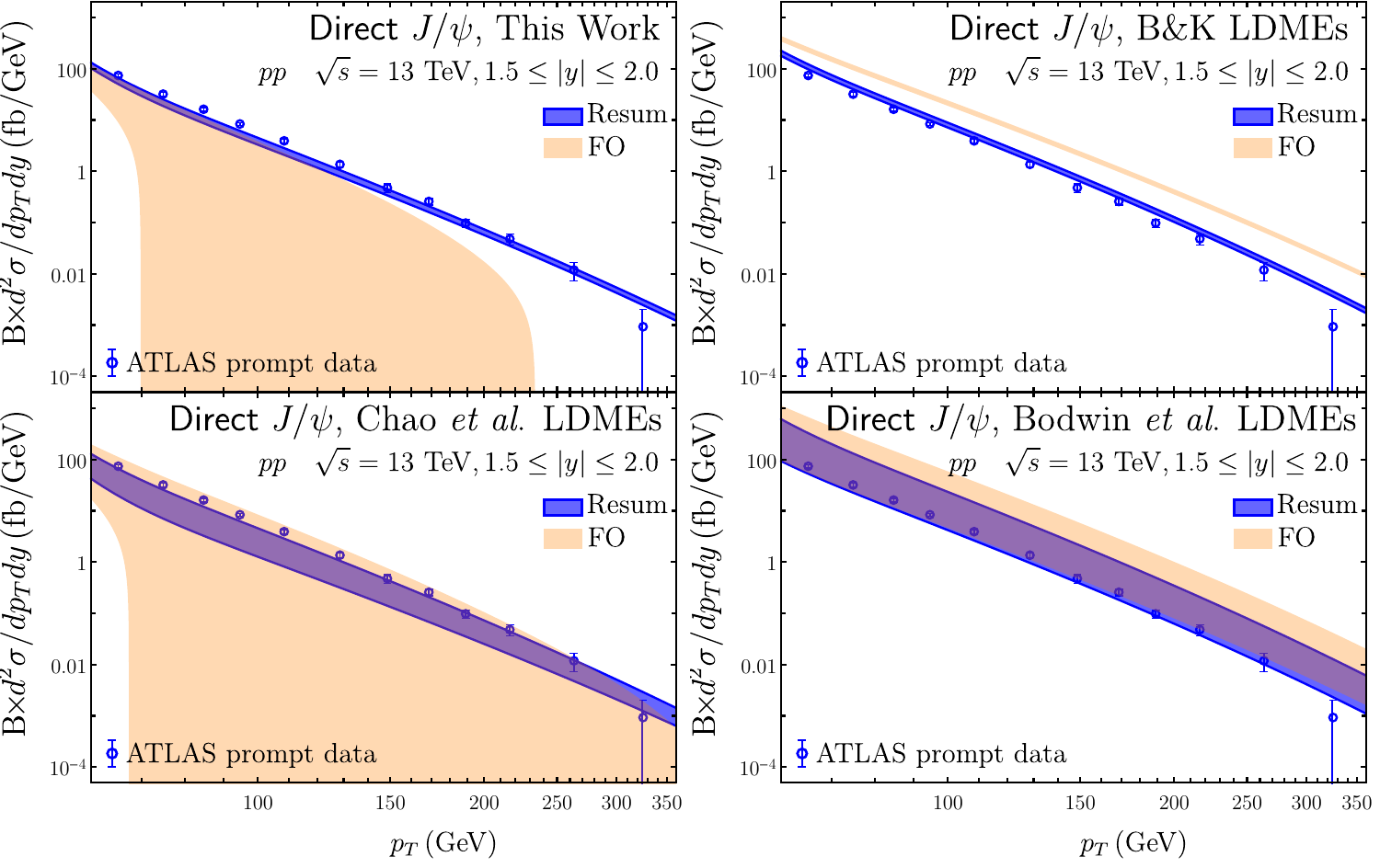}%
\caption{\label{fig:otherMEs}
Direct $J/\psi$ production rates from $pp$ collisions at $\sqrt{s}=13$~TeV
for the rapidity range $1.5 \le |y| \le 2.0$ computed by using LDMEs adopted in
this work, compared to ones 
from Refs.~\cite{Butenschoen:2011yh} (B\&K), 
\cite{Han:2014jya} (Chao {\it et al.}), 
and \cite{Bodwin:2014gia} (Bodwin {\it et al.}). 
ATLAS data for prompt $J/\psi$ cross section are shown for comparison.
}
\end{figure}

In Fig.~\ref{fig:otherMEs}, we show results for direct $J/\psi$ cross sections
computed from other LDME determinations in Refs.~\cite{Butenschoen:2011yh,
Han:2014jya, Bodwin:2014gia}. 
Note that the LDMEs in Ref.~\cite{Butenschoen:2011yh} (B\&K) lead to transverse
polarization, which is incompatible with measurements~\cite{CMS:2013gbz}. 
It is known that the LDMEs from Ref.~\cite{Bodwin:2014gia} 
(Bodwin {\it et al.}) are
incompatible with $\eta_c$ data~\cite{LHCb:2014oii, LHCb:2019zaj}. 
It is important to note that, regardless of the choice of LDMEs, 
in a FO calculation the $\chi_{cJ}$ cross sections always turn negative, 
so that any solid prediction of prompt $J/\psi$ production rate at large $p_T$ 
requires resummation of threshold logarithms.

{\it Conclusions---}The resummation of threshold double logarithms we computed
in this Letter substantially improves the NRQCD description of charmonium
production rates at large transverse momentum. This resolves the catastrophic
failure of fixed-order perturbation theory where large-$p_T$ cross sections can
turn unphysically negative. Resummation of threshold logarithms may also be
important in describing measurements involving kinematical cuts near the
boundary of phase space such as photoproduction rates~\cite{ZEUS:2002src,
H1:2002voc, H1:2010udv} and quarkonium in jet~\cite{Baumgart:2014upa,
LHCb:2017llq, Kang:2017yde, Bain:2017wvk, CMS:2019ebt}. 
In the case of bottomonium, we expect that the effect of threshold logarithms 
will become important if the transverse momentum exceeds 1~TeV, based on the
LDME determinations in Refs.~\cite{Han:2014kxa, Brambilla:2021abf, Brambilla:2022ayc}.
It would be interesting to improve the accuracy of
the resummation to single logarithmic level, which requires calculation of the
soft functions to single-pole accuracy. The resummation may also be extended to
NLP fragmentation contributions, which become important for lower $p_T$.

\begin{acknowledgments}
{\it Acknowledgments---}We thank Geoffrey Bodwin for fruitful discussions on resummation of threshold 
logarithms. 
The work of H.~S.~C. and J.~L. is supported by  the National Research
Foundation of Korea (NRF) Grant funded by the Korea government (MSIT) under
Contract No. NRF2020R1A2C3009918. H.~S.~C. also acknowledges support by the
Basic Science Research Program through the National Research Foundation of
Korea (NRF) funded by the Ministry of Education (Grant No. RS-2023-00248313)
and a Korea University grant. 

H.~S.~C., U-R.~K., and J.~L. 
contributed equally to this work.
\end{acknowledgments}

\bibliography{threshold_short.bib}

\end{document}